\documentclass[conference]{IEEEtran}

\usepackage{graphicx}
\usepackage{amsmath}
\usepackage{cite}
\usepackage{url}
\usepackage[draft]{hyperref}
\usepackage{graphicx}
\usepackage{array}
\usepackage{booktabs}
\usepackage{multirow}
\usepackage{graphicx}
\usepackage{subcaption}
\usepackage{cleveref}
\usepackage{array}

\usepackage{listings}
\usepackage{xcolor}

\title{Understanding BBRv3 Performance in AQM-Enabled WiFi Networks}

\author{\IEEEauthorblockN{Shyam Kumar Shrestha, Jonathan Kua and Shiva Raj Pokhrel}
\IEEEauthorblockA{School of Information Technology \\  Deakin University, Geelong, Australia \\
\{shyam.shrestha, jonathan.kua, shiva.pokhrel\}@deakin.edu.au}
}

\begin{document}

\maketitle

\begin{abstract}

We present a modular experimental testbed and lightweight visualization tool for evaluating TCP congestion control performance in wireless networks. We compare Google's latest Bottleneck Bandwidth and Round-trip time version 3 (BBRv3) algorithm with loss-based CUBIC under varying Active Queue Management (AQM) schemes, namely PFIFO, FQ-CoDel, and CAKE, on a Wi-Fi link using a commercial MikroTik router. Our real-time dashboard visualizes metrics such as throughput, latency, and fairness across competing flows. Results show that BBRv3 significantly improves fairness and convergence under AQM, especially with FQ-CoDel. Our visualization tool and modular testbed provide a practical foundation for evaluating next-generation TCP variants in real-world AQM-enabled home wireless networks.

\end{abstract}


\section{Introduction} \label{introduction}

The widespread use of Wi-Fi networks for data transmission, particularly in residential and enterprise environments, introduces significant complexities in Transmission Control Protocol (TCP) performance. Key challenges include achieving higher resource utilization, maintaining minimum latency maximizing throughput~\cite{zeynali2024promises,shrestha2024fairness}, and minimizing packet loss and retransmissions. In response to these ongoing issues, researchers and developers have proposed improvements to TCP's inherent congestion control mechanisms. Over time, Congestion Control Algorithms (CCAs) have evolved to address the performance demands of increasingly heterogeneous and dynamic network environments.

A recent milestone in congestion control evolution is the development of Bottleneck Bandwidth and Round-trip time version 3 (BBRv3)~\cite{cardwell2023bbrv3}, developed and deployed on Google platforms since 2023. Building on lessons learnt from BBRv1~\cite{cardwell2017bbr} and BBRv2\cite{cardwell2018bbr}, BBRv3 enhances fairness with competing flows, especially in shallow-buffer and dynamic network conditions. It addresses key limitations in earlier versions by refining pacing, congestion window control, and responsiveness to loss signals. Notably, BBRv3 improves intra-protocol fairness and ensures more consistent bandwidth probing, helping flows converge more equitably in both low- and high-buffer environments.

In this demo, we present a modular experimental testbed with a visualization tool to highlight BBRv3’s improved performance in contention-limited Wi-Fi setups, when competing with loss-based CCAs such as TCP CUBIC~\cite{ha2008cubic}. Our work serves as a practical tool for evaluating network performance for new and emerging CCAs in a home Wi-Fi networks~\cite{9468697, 9344582}.

Our core observation revolves around whether BBRv3 provides demonstrable advantages and is indeed better suited for a simple home Wi-Fi environment than widely-deployed loss-based CCAs, such as TCP CUBIC. Furthermore, we evaluate their actual performance under an emulated 10 Mbps bottleneck, a common constraint in many home internet connections. We test these TCP algorithms with different queuing disciplines (qdisc), ranging from the fundamental Packet First-In First-Out (PFIFO) to more sophisticated approaches such as FlowQueue-Controlled Delay (FQ-CoDel)~\cite{hoeiland2018flow} and Common Applications Kept Enhanced (CAKE)~\cite{hoiland2018piece}. This multi-faceted analysis will provide deeper insights into how each algorithm manages congestion and maintains data flow stability across diverse queuing strategies in practical home settings and Internet of Things (IoT) environments~\cite{kua2017using,satish2024active}.

\section{Data Visualization} \label{visualization}

To visualize the dynamics of competing TCP flows in real-time, we built a lightweight Python-based dashboard using Plotly's \textit{Dash\footnote{\url{https://dash.plotly.com/}}} framework.

This dashboard enables clear, customizable visual comparisons of transport-layer performance, especially useful for showcasing how modern congestion control mechanisms such as BBRv3 react to congestion, compete for bandwidth, or adapt to varying wireless link dynamics.

\begin{figure}[htbp]
    \centering
    \begin{subfigure}[b]{0.46\textwidth}
        \centering
        \includegraphics[height=3.7cm, keepaspectratio]{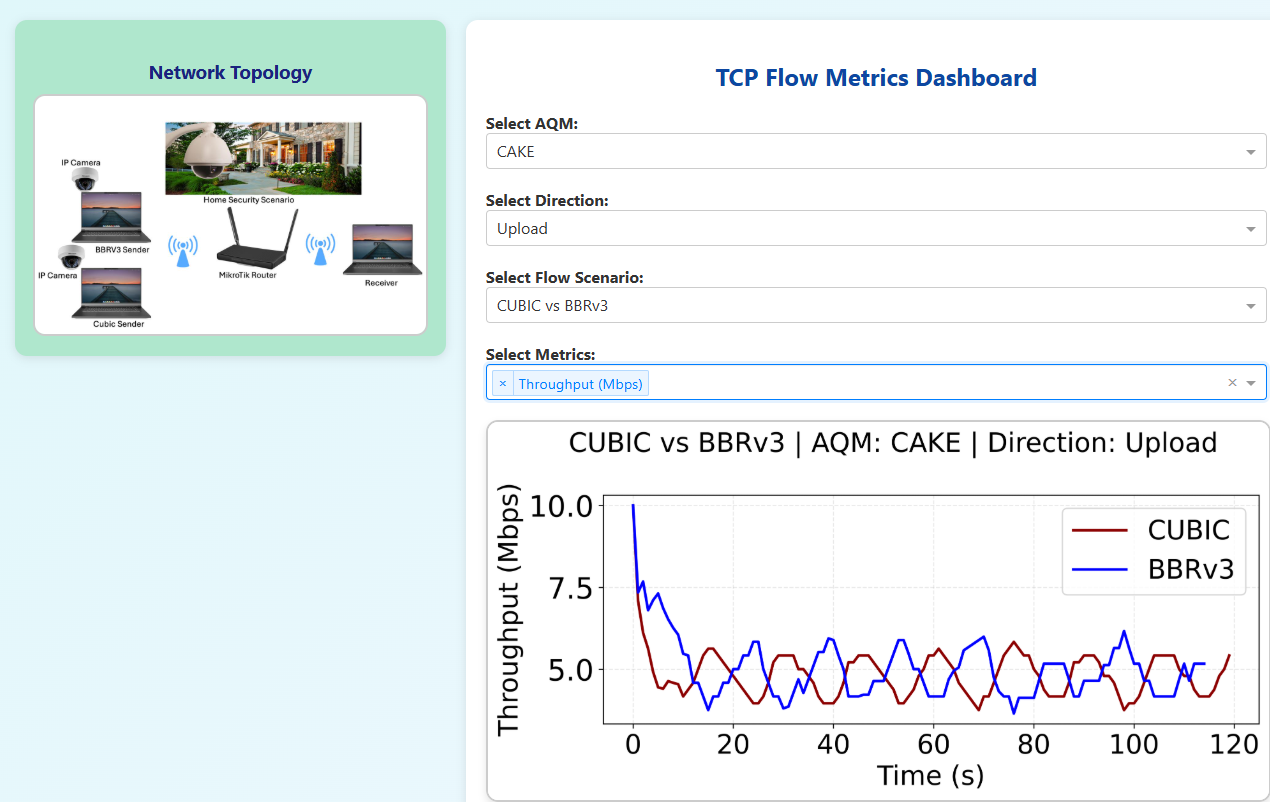}
        \caption{Throughput comparison between BBRv3 and CUBIC in competing upload scenario over CAKE AQM.}
        \label{fig:viz_thrupt}
    \end{subfigure}
    \hfill
    \begin{subfigure}[b]{0.46\textwidth}
        \centering
        \includegraphics[height=3.7cm, keepaspectratio]{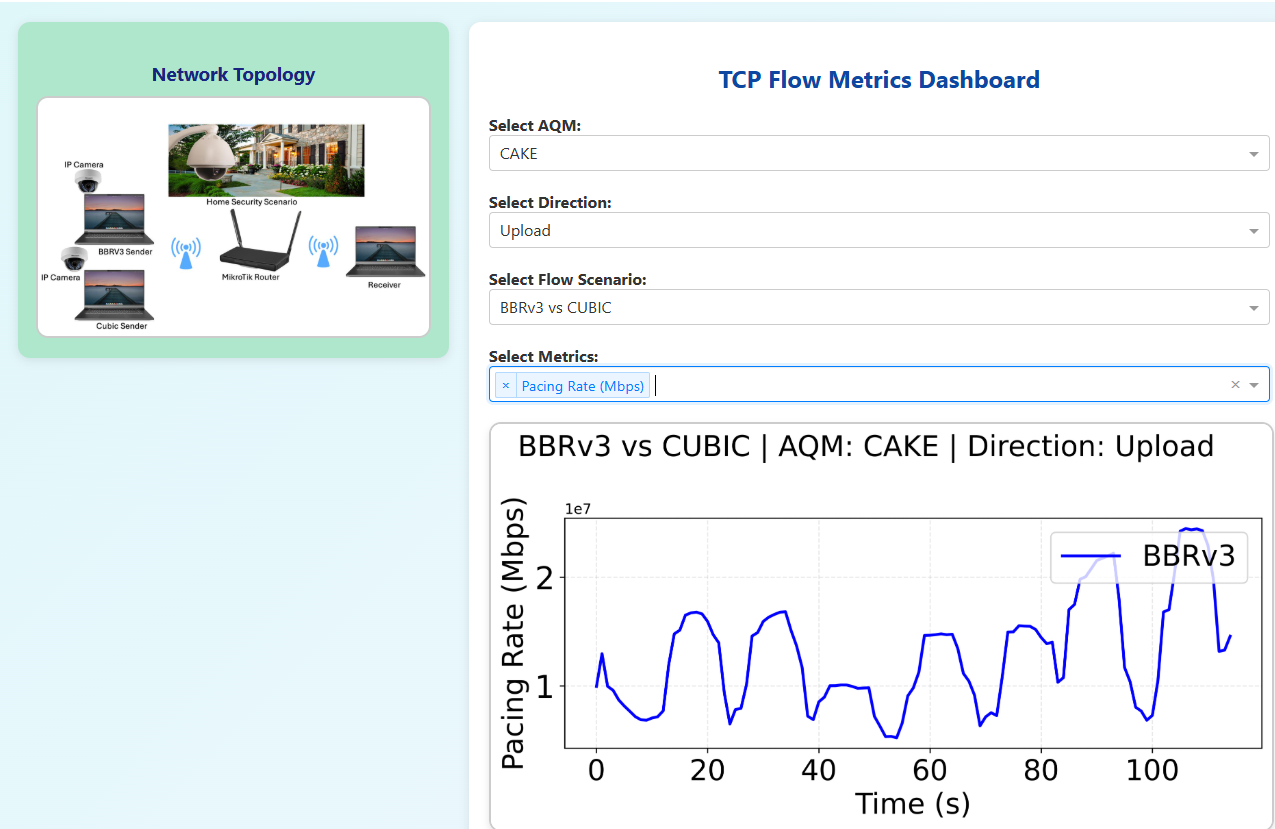}
        \caption{Pacing behavior of BBRv3 when competing with CUBIC over CAKE AQM.}
        \label{fig:pacing_rate}
    \end{subfigure}
    \hfill
    \begin{subfigure}[b]{0.46\textwidth}
        \centering
        \includegraphics[height=3.7cm, keepaspectratio]{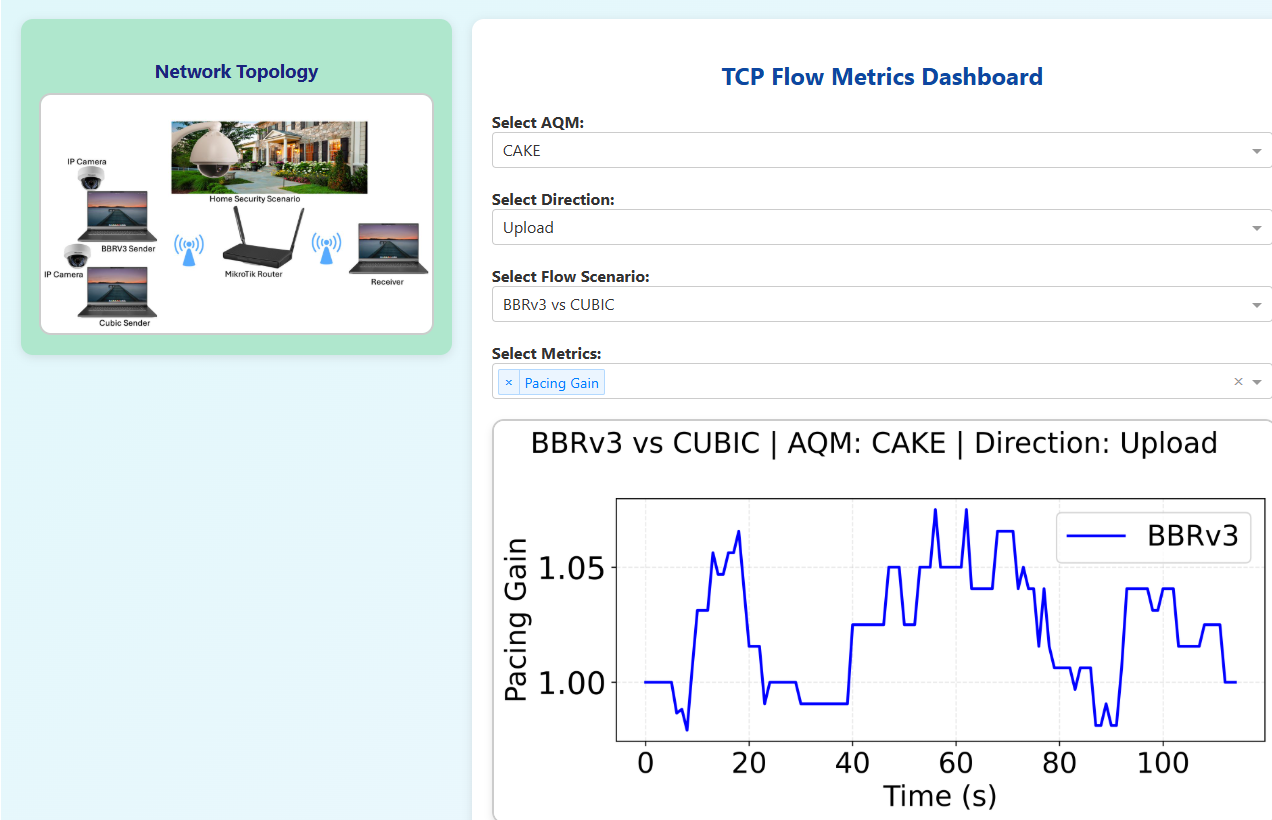}
       \caption{Pacing gain dynamics of BBRv3 when competing with CUBIC over CAKE AQM.}
        \label{fig:pacing_gain}
    \end{subfigure}

    \caption{Dashboard of our data analysis and visualization tool.}
    \label{fig:vizpic}
\end{figure}

Fig.~\ref{fig:vizpic} provides a representative example of the interactive visualization tool developed for analyzing TCP congestion control algorithm performance within our experimental Wi-Fi testbed. This tool facilitates the detailed exploration of key TCP metrics, including throughput, RTT, jitter, and congestion window (\textit{cwnd}), retransmissions for competing flows (e.g., CUBIC vs. BBRv1, BBRv2, BBRv3) under different AQM schemes (PFIFO, FQ-CoDel, CAKE). BBRv3-specific parameters such as pacing gain and pacing rate are also included. 

For instance, Fig.~\ref{fig:viz_thrupt} illustrates throughput performance when BBRv3 competes with CUBIC, showing clear disparities under PFIFO. Fig.~\ref{fig:pacing_rate} highlights BBRv3’s pacing rate adaptation in such conditions, while Fig.~\ref{fig:pacing_gain} depicts its pacing gain across phases and its link to rate changes and throughput variability.

By offering an interactive selection of TCP CCAs, AQMs, and performance metrics, the dashboard enables real-time comparison and deep dives into the dynamic behavior of these algorithms across a bottleneck with configurable bandwidth and latency requirements. Furthermore, the modular design of our experimental setup and data analysis/visualization framework allows for ready adaptation to assess new CCAs, diverse network topologies, or varying bottleneck capacities in future research. Visual representation and analysis of CCAs behaviors are crucial for understanding how they interact, adapt and perform in practical Wi-Fi environments.

\section{Modular Experimental Testbed} \label{methodology}

This section outlines our modular experimental testbed setup, which is used in conjunction with our visualization framework for real-time evaluation of CCAs performance in Wi-Fi networks.

\begin{figure}[h]
\centering
\includegraphics[scale=0.15]{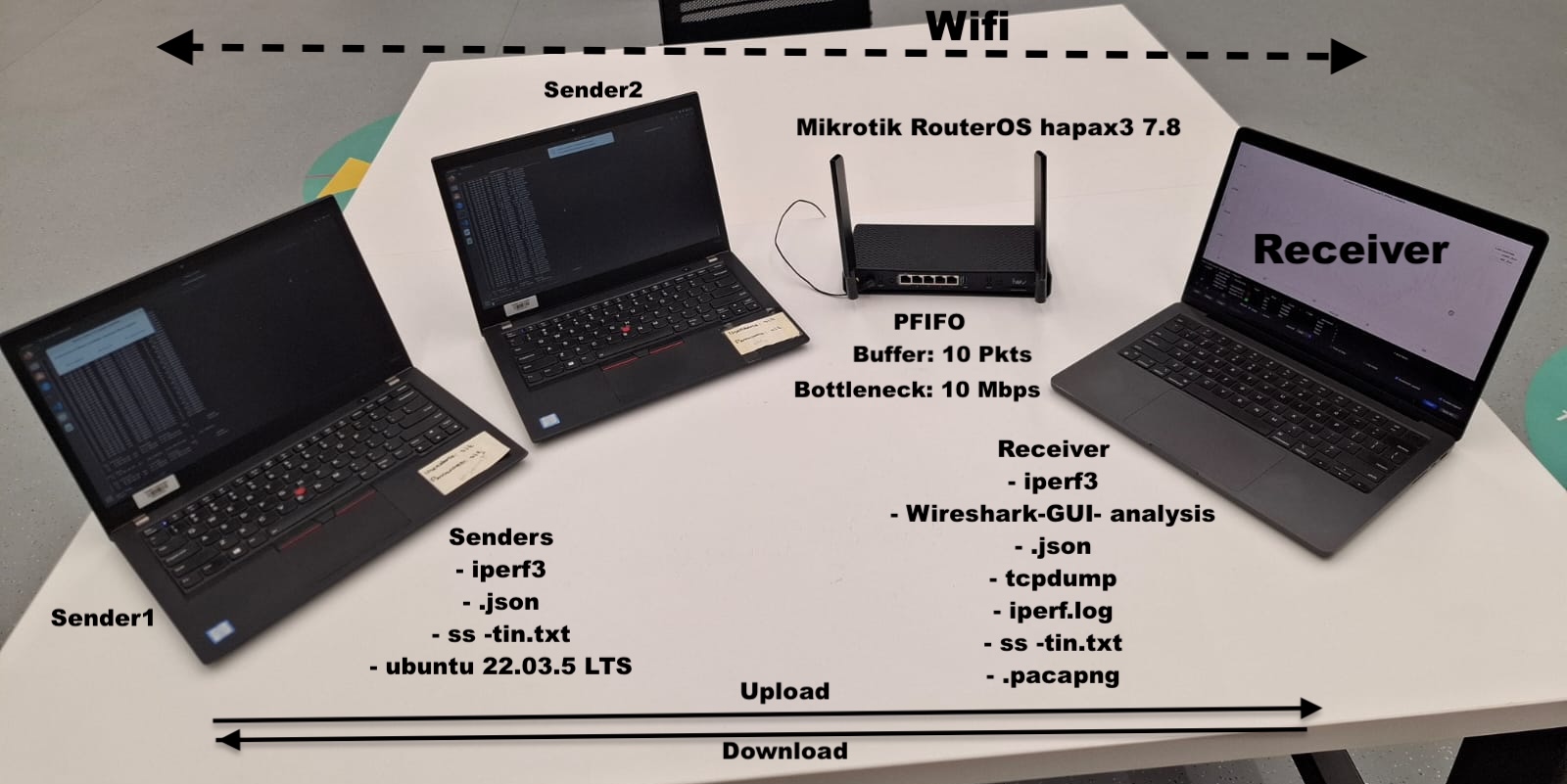}
\caption{Modular experimental testbed for evaluating TCP BBRv3 under controlled AQM-enabled Wi-Fi bottlenecks.}
\label{fig:testbed}
\end{figure}

Fig.~\ref{fig:testbed} illustrates a customized wireless testbed architecture designed to evaluate CCAs under realistic Wi-Fi conditions, encompassing both homogeneous and heterogeneous scenarios. At the core is a MikroTik router (hap ax3, RouterOS v7.8; SSID: Mikrotik2), configured with two independent wireless interfaces and distinct subnets to emulate upload and download bottlenecks separately. 

The sender nodes (Ubuntu 22.04.5 LTS) are connected to `wifi1`, which is bridged with the Ethernet interface (`Bridge-LAN`) and assigned to the 192.168.10.0/24 subnet. The receiver node connects to a separate Wi-Fi interface, `wifi2`, operating within the 192.168.20.0/24 subnet. This network segmentation facilitates precise bandwidth shaping and queuing control through MikroTik’s Queue Tree and Firewall Mangle rules. It ensures that queue management policies can be applied independently in each direction, enabling accurate emulation of congestion, contention, and queuing behaviors for robust evaluation of CCAs under Wi-Fi impairments.

To emulate constrained wireless bandwidth, MikroTik’s Simple Queue and Queue Tree mechanisms are configured to limit throughput to 10 Mbps in both the upload and download directions. AQM schemes are configured using a strict PFIFO policy with a fixed buffer size of 50 packets, alongside modern AQM schemes, such as FQ-CoDel and CAKE, to enable a comparative analysis of PFIFO vs AQM schemes. This setup allows the study of congestion buildup, tail-drop behavior, and fairness among different CCAs, including BBR variants and CUBIC, under contention-limited Wi-Fi conditions.

TCP bulk transfer traffic is generated using \textit{iperf3} (v3.9), while real-time socket-level statistics are collected via \textit{ss -tin}. Packet-level traces captured using \textit{tshark} (v4.4.6). The raw outputs are parsed into structured JSON logs using the \textit{cJSON} library (v1.7.3). Additional client-side metrics such as throughput, \textit{cwnd}, RTT, and jitter are simultaneously logged.

The generated experimental datasets support modular and reproducible evaluations of TCP behaviors under Wi-Fi-specific conditions, such as channel contention, bursty losses, and queue-induced bottlenecks. We have made our experimental scripts and detailed router configuration/host setup instructions available via our GitHub repository.\footnote{\url{https://github.com/Shyamesh/LCN_Demo}}.

\begin{figure}[h]
\centering
\includegraphics[scale=0.3]{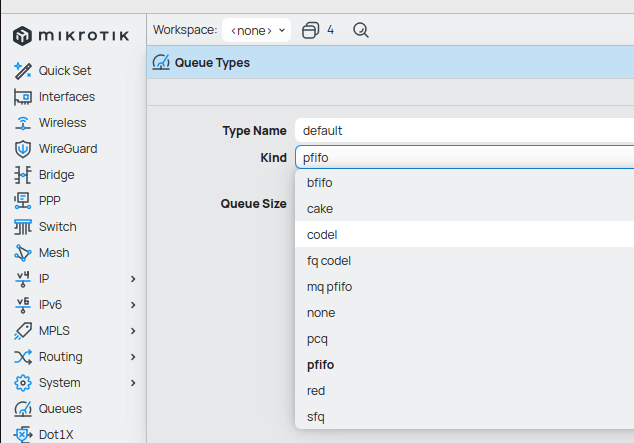}
\caption{AQM configurations in MikroTik RouterOS (hAP ax3, version 7.8), used in our experimental testbed.}
\label{fig:aqm}
\end{figure}

Fig.~\ref{fig:aqm} shows the six AQM mechanisms available in MikroTik RouterOS. We selected FQ-CoDel and CAKE for benchmarking, using PFIFO as a comparison baseline. FQ-CoDel is chosen to evaluate whether competing flows governed by different CCAs receive a fair share of bandwidth. CAKE was selected to assess its performance under constrained bandwidth conditions, as it is designed to perform well in such environments in the real-world.

Table~\ref{table1} summarizes the experimental components, their configurations, and their respective roles in the testbed.

\begin{table}[h]
\centering
\normalsize
\caption{Experimental testbed parameters and configurations.}
\label{table1}
\resizebox{\columnwidth}{!}{%
\begin{tabular}{|p{2.5cm}|p{3.5cm}|p{3.5cm}|}
\hline
\textbf{Component} & \textbf{Configuration} & \textbf{Function} \\
\hline
Bottleneck router & MikroTik RouterOS (hAP ax3,
version 7.8) & Wireless network with in-built AQM \\
\hline
IP address space & 192.168.10.0/24 and 192.168.20.0/24 & Assigns static IPs to sender and receiver nodes  \\
\hline
Wireless link bandwidth & Nominal 10 Mbps & Emulates real-world wireless bandwidth \\
\hline
Queue management policies & PFIFO, FQ-CoDel, and CAKE on MikroTik router & Evaluates TCP performance under different queuing schemes \\
\hline
Server Ethernet interface & Port 2 disabled during data acquisition & Isolates analysis to wireless communication \\
\hline
TCP traffic generation & \textit{iperf3} v3.9 & Creates representative TCP traffic flows \\
\hline
Data collection and analysis & \textit{tshark} v4.4.6 (CLI of Wireshark) & Captures and analyses network traffic in \textit{.pcapng} \\
\hline
Data formatting & \textit{cJSON} v1.7.3 & Structured output for TCP metrics \\
\hline
Client-side data acquisition & \textit{ss -tin} command & Collects system-level socket statistics \\
\hline
Operating system & Ubuntu 22.04.5 LTS & Uniform operating system across all nodes \\
\hline
\end{tabular}%
}
\end{table}

\section{Evaluation and Analysis} \label{analysis}

\begin{figure}[htbp]
    \centering
    \begin{subfigure}[b]{0.15\textwidth}
        \centering
        \includegraphics[width=\textwidth]{CUBIC_vs_bbrv3_pfifo_up.png}
        \caption{PFIFO}
        \label{fig:pfifo_up}
    \end{subfigure}
    \hfill
    \begin{subfigure}[b]{0.15\textwidth}
        \centering
        \includegraphics[width=\textwidth]{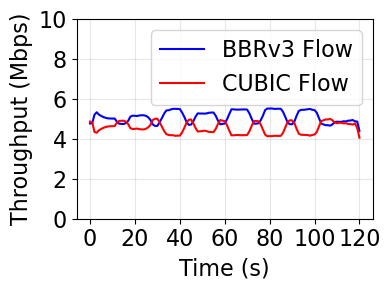}
        \caption{FQ-CoDel}
        \label{fig:Fq-codel_up}
    \end{subfigure}
    \hfill
    \begin{subfigure}[b]{0.15\textwidth}
        \centering
        \includegraphics[width=\textwidth]{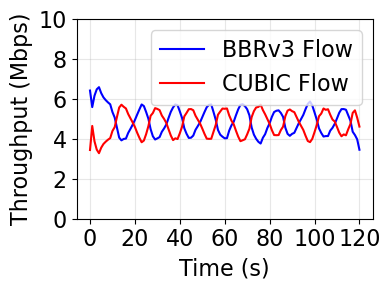}
        \caption{CAKE}
        \label{fig:CAKE_up}
    \end{subfigure}

    \caption{Performance of BBRv3 when competing with CUBIC under (a) PFIFO, (b) FQ-CoDel, and (c) CAKE AQM schemes in the upload direction over Wi-Fi.}
    \label{fig:bbr_comparison_up}
\end{figure}

\begin{figure}[htbp]
    \centering
    \begin{subfigure}[b]{0.15\textwidth}
        \centering
        \includegraphics[width=\textwidth]{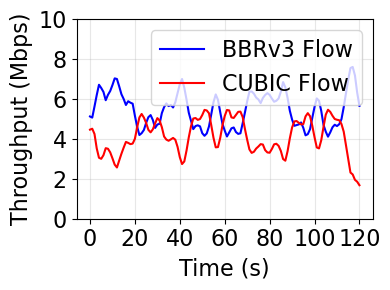}
        \caption{PFIFO}
        \label{fig:pfifo_dw}
    \end{subfigure}
    \hfill
    \begin{subfigure}[b]{0.15\textwidth}
        \centering
        \includegraphics[width=\textwidth]{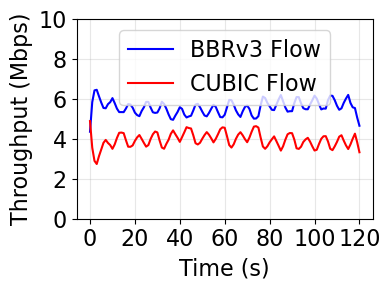}
        \caption{FQ-Codel}
        \label{fig:Fq-codel_dw}
    \end{subfigure}
    \hfill
    \begin{subfigure}[b]{0.15\textwidth}
        \centering
        \includegraphics[width=\textwidth]{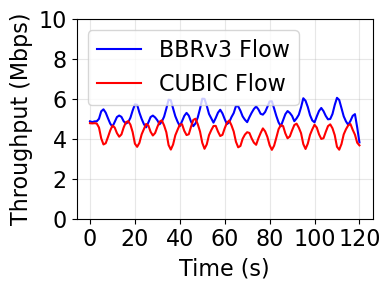}
        \caption{CAKE}
        \label{fig:CAKE_dw}
    \end{subfigure}

    \caption{Performance of BBRv3 when competing with CUBIC under (a) PFIFO, (b) FQ-CoDel, and (c) CAKE AQM schemes in the download direction over Wi-Fi.}
    \label{fig:bbr_comparison_download}
\end{figure}

Our evaluation reveals the distinct operational characteristics of three queuing disciplines PFIFO, FQ-CoDel, and CAKE under inter-TCP variant competition between CUBIC and BBRv3 within Wi-Fi environments. The analysis encompasses both upload (multiple sender send packets to a single receiver) and download (one sender sends packets to multiple receivers) scenarios, where the competing flows share a constrained 10 Mbps bottleneck over a 120-second interval, as illustrated in Figs.~\ref{fig:bbr_comparison_up} and~\ref{fig:bbr_comparison_download}.

Under the default PFIFO queuing mechanism (Figs.~\ref{fig:pfifo_up} and~\ref{fig:pfifo_dw}), BBRv3 exhibits asymmetric performance. It remains competitive and even dominant in the download case (Fig.~\ref{fig:pfifo_dw}), but suffers significant performance degradation in the upload scenario (Fig.~\ref{fig:pfifo_up}) when coexisting with CUBIC flows. This disparity arises from PFIFO’s shared buffer architecture, which disproportionately favors CUBIC’s aggressive, loss-driven congestion probing. As CUBIC fills the queue, it introduces additional queuing delay, which inflates RTT. BBRv3, being sensitive to RTT increases, interprets the elevated delay as a signal of congestion and reacts conservatively by reducing its sending rate, ultimately resulting in suppressed throughput in the upload direction.

In contrast, Figs.~\ref{fig:Fq-codel_up} and \ref{fig:Fq-codel_dw} demonstrates the effectiveness of FQ-CoDel in improving flow isolation and fairness particularly in upload scenarios where multiple senders transmit data to a single receiver. Under this queuing discipline, BBRv3 consistently dominates CUBIC, achieving significantly higher throughput. This dominance is more evident in the download direction (Fig.~\ref{fig:Fq-codel_dw}). In both upload and download cases, BBRv3 benefits from FQ-CoDel’s capability to maintain flow pacing and protect delay-sensitive flows from aggressive, loss-based variants such as CUBIC. The per-flow queuing and active queue management of FQ-CoDel mitigate buffer contention, enabling BBRv3 to fully utilize its rate-based probing strategy. As a result, bandwidth sharing becomes not only more efficient but also more favorable to BBRv3.

Similarly, Figs.~\ref{fig:CAKE_up} and \ref{fig:CAKE_dw} illustrate that CAKE provides robust management of competing flows, delivering consistent performance across both upload and download scenarios with only minor variations. Under CAKE, BBRv3 exhibits more harmonious coexistence with its loss-based counterpart, CUBIC, showing less aggressive dominance than under FQ-CoDel. However, in the download direction (Fig.~\ref{fig:CAKE_dw}), BBRv3 still secures higher throughput, even at the cost of increased retransmissions.

This favorable outcome for BBRv3 can be attributed to CAKE’s flow-aware active queue management and deficit-based fair scheduling, which preserve pacing and reduce queue buildup. Since BBRv3 relies on RTT signals and pacing to adapt its sending rate, CAKE's ability to isolate flows and control delay helps BBRv3 avoid overreacting to transient congestion. Consequently, BBRv3 maintains high sending rates and benefits from reduced interference by loss-based flows, especially in receiver-driven traffic where CAKE’s fairness policies are more impactful.

In summary, this evaluation highlights the critical role of queuing discipline in shaping congestion control and fairness. FIFO-based queues such as PFIFO hinder BBRv3, especially in uploads, by increasing contention and delay. In contrast, fair queuing schemes such as FQ-CoDel and CAKE enhance performance by isolating flows and reducing latency, allowing BBRv3 to fully exploit its pacing and RTT-aware design. These results underscore the need for smarter queue management to support diverse TCP variants, ensure fair coexistence in real-world deployments.

\begin{figure}[htbp]
    \centering
    \begin{subfigure}[b]{0.15\textwidth}
        \centering
        \includegraphics[width=\textwidth]{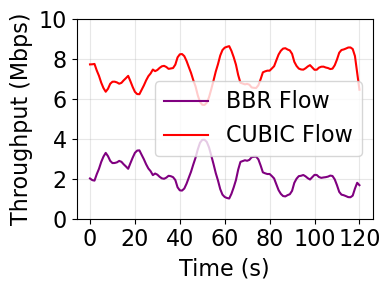}
        \caption{BBRv1}
        \label{fig:bbr1_CUBICvsbbr}
    \end{subfigure}
    \hfill
    \begin{subfigure}[b]{0.15\textwidth}
        \centering
        \includegraphics[width=\textwidth]{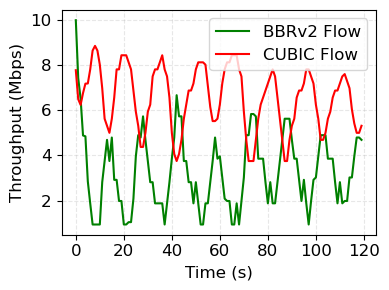}
        \caption{BBRv2}
        \label{fig:bbr2_CUBICvsbbr}
    \end{subfigure}
    \hfill
    \begin{subfigure}[b]{0.15\textwidth}
        \centering
        \includegraphics[width=\textwidth]{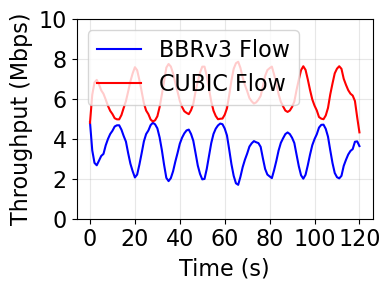}
        \caption{BBRv3}
        \label{fig:bbr3_CUBICvsbbr}
    \end{subfigure}

    \caption{Throughput comparison of different BBR versions: (a) BBRv1, (b) BBRv2, and (c) BBRv3 when competing with CUBIC under the PFIFO over WiFi.}
    \label{fig:bbr_versions_over_fifo}
\end{figure}

In addition, we evaluate the performance of BBR variants BBRv1, BBRv2, and BBRv3 when competing against the loss-based TCP CUBIC under the default PFIFO queuing discipline. Our objective is to examine whether newer BBR versions show progress in key areas such as fairness and convergence. As shown in Fig.~\ref{fig:bbr1_CUBICvsbbr}, BBRv1 performs poorly, experiencing near-starvation throughout the 120-second experiment. Despite this, BBRv1 aggressively probes for bandwidth without sufficiently reacting to queuing delays or packet loss, often leading to starvation of competing flows indicating a lack of fairness in shared network environments.

BBRv2 (Fig.~\ref{fig:bbr2_CUBICvsbbr}) shows improvement, reaching convergence with CUBIC around the 40-second mark, though with significant throughput fluctuations. BBRv2 introduces a more conservative bandwidth probing strategy and attempts to better share bandwidth by reacting to queuing signals. While this version shows improved coexistence with CUBIC reducing its own throughput to avoid overwhelming the bottleneck it suffers from significant throughput fluctuations and slower convergence due to its cautious probing and recovery behavior.

In contrast, BBRv3 (Fig.~\ref{fig:bbr3_CUBICvsbbr}) demonstrates substantial improvements in both stability and coexistence. Although CUBIC still achieves slightly higher throughput, BBRv3 maintains smoother throughput curves and avoids starvation. These improvements stem from BBRv3’s refined gain cycling, enhanced RTT tracking, and queue buildup detection, which allow it to balance aggressiveness with responsiveness more effectively than its predecessors.

Overall, the results indicate that BBRv3 incorporates meaningful enhancements that support fairer and faster adaptation in mixed TCP environments especially under simple queuing disciplines such as PFIFO where there are no in-network flow isolation capabilities.

\section{Conclusions and Future Work} \label{conclusion}

This demo showcases a modular experimental testbed and data visualization framework for assessing new and emerging TCP CCAs in Wi-Fi environments. Our observations highlight that BBRv3 shows noticeable improvement in fairness and convergence compared to its predecessors, especially under AQM schemes such as FQ-CoDel and CAKE.

Future work will include extending our testbed and visualization tool to include wider range of experimental scenarios with tunable in-network configurations/parameters and Machine Learning-driven approaches. Use cases with real-world networking devices, such as those for smart home automation and enterprise-level network telemetry, will also be an important feature of our work.

\bibliographystyle{IEEEtran}
\bibliography{refs}

\end{document}